# Enhanced electron correlations, local moments, and Curie temperature in strained MnAs nanocrystals embedded in GaAs

M. Moreno,[a,*] J. I. Cerdá,[a] K. H. Ploog,[b] and K. Horn[c]

[a] *Instituto de Ciencia de Materiales de Madrid (CSIC), Cantoblanco, E-28049 Madrid, Spain*

[b] *Paul-Drude-Institut für Festkörperelektronik, Hausvogteiplatz 5-7, D-10117 Berlin, Germany*

[c] *Fritz-Haber-Institut der Max-Planck-Gesellschaft, Faradayweg 4-6, D-14195 Berlin, Germany*

(Dated: August 4, 2011)

## ABSTRACT

We have studied the electronic structure of hexagonal MnAs, as epitaxial continuous film on GaAs(001) and as nanocrystals embedded in GaAs, by Mn 2*p* core-level photoemission spectroscopy. Configuration-interaction analyses based on a cluster model show that the ground state of the embedded MnAs nanocrystals is dominated by a $d^5$ configuration that maximizes the local Mn moment. Nanoscaling and strain significantly alter the properties of MnAs. Internal strain in the nanocrystals results in reduced *p-d* hybridization and enhanced ionic character of the Mn-As bonding interactions. The spatial confinement and reduced *p-d* hybridization in the nanocrystals lead to enhanced *d*-electron localization, triggering *d-d* electron correlations and enhancing local Mn moments. These changes in the electronic structure of MnAs have an advantageous effect on the Curie temperature of the nanocrystals, which is measured to be remarkably higher than that of bulk MnAs.

---

[*] E-mail: mmoreno@icmm.csic.es

# I – INTRODUCTION

The first generation of spintronic devices has used metallic magnetic materials with enormous success for data-storage applications.[1] The next generation, under development, further aims at monolithic integration of information storage and processing tasks, therefore it is planned to be semiconductor based.[2-4] Whereas the first generation has used transition metals (TMs) in elemental or alloyed form as magnetic materials, the next generation is envisaged to use TM coordination compounds, consisting of "*d*" cations (TM atoms) and "*sp*" anions (group IV to VI atoms), in order to achieve higher structural and electronic compatibility with the semiconductor materials used in conventional electronics. The different atomic species of the TM compound play different roles in the magnetic and spin-transport properties. On the one hand, the TM cations contribute robust strongly localized "inducing" magnetic moments, formed primarily because of the intra-atomic exchange interaction. On the other hand, the sublattice of nonmagnetic anions may develop, through (*sp*)-*d* hybridization, "induced" moments with delocalized character. Although magnetism arises from the TM cations, the possibility of spin transport (the essential aim of semiconductor spintronics) strongly depends on the moments induced in the anion sublattice. The characteristics of the (*sp*)-*d* hybridization determine whether moments are induced in the intrinsically nonmagnetic anion sublattice, and how delocalized the induced moments are.

The recent use of nonequilibrium methods (low temperature) in molecular-beam epitaxy to fabricate hybrid ferromagnet-semiconductor structures has led to a breakthrough for the advancement of semiconductor spintronics. Many of the ferromagnetic materials grown up to now incorporate Mn, as its half-filled *d* atomic shell contributes large "inducing" magnetic moments. Mn has been successfully incorporated into GaAs, as $Ga_{1-x}Mn_xAs$ ($x \lesssim 0.1$) diluted magnetic semiconductor, maintaining the GaAs zinc-blende (cubic) structure,[5] and ferromagnetic hexagonal MnAs films have been grown on GaAs and Si substrates, with high structural perfection and atomically sharp interfaces.[6] Besides, granular GaAs:MnAs materials, consisting of MnAs nanocrystals embedded in an essentially Mn-free semiconducting GaAs matrix, have been prepared by high-temperature annealing of low-temperature grown $Ga_{1-x}Mn_xAs$.[7]

The different crystal structure and chemical composition of MnAs and $Ga_{1-x}Mn_xAs$ are responsible for the different electronic structure of these compounds, and this in turn is responsible for the different magnetic and transport properties. Whereas optimally prepared $Ga_{1-x}Mn_xAs$ ($x \sim 0.07$) is a highly *p*-type-doped semiconductor, ferromagnetic up to ~173 K (Ref. [8]), MnAs exhibits metallic conductivity and is ferromagnetic up to ~313 K (Ref. [9]). In both materials, a magnetic moment is induced in the As sublattice, with opposite sign to the local Mn moments. At present, optimally prepared $Ga_{1-x}Mn_xAs$ ($x \sim 0.07$) is, among diluted magnetic semiconductors, the material exhibiting the best spintronic performance. However, its Curie temperature is still too low for practical applications. On the other hand, MnAs is ferromagnetic at room temperature, but its metallic conductivity limits efficient spin injection into semiconductors because of the resistivity mismatch. A material exhibiting both room-temperature ferromagnetism, as MnAs, and semiconductor resistivity, as $Ga_{1-x}Mn_xAs$, would be ideal for semiconductor spintronics. In order to see how these requirements could be fulfilled, it is important to fully understand the Mn-Mn and Mn-As bonding interactions in $Ga_{1-x}Mn_xAs$ and MnAs related materials, and how these interactions affect ferromagnetism stability and electrical conduction properties.

In this paper, we present a comparative study of the electronic structures of nanocluster and film morphologies of hexagonal MnAs, by photoemission spectroscopy (PES). The Mn 2*p* photoemission lineshape from MnAs is analyzed in detail. Brief photoemission studies were previously reported on zinc-blende-type MnAs nanodots on passivated GaAs surfaces.[10,11] Photoemission spectroscopy in combination with configuration-interaction (CI) cluster-model analyses has been a key experimental tool to study the electronic properties of Mn impurities in diluted magnetic semiconductors.[12-14] Here, we present a CI cluster-model analysis of the Mn 2$p$ photoemission lineshape from hexagonal MnAs nanoclusters embedded in GaAs, providing insight into the effect of nanoscaling and internal strain on the electronic and magnetic properties of MnAs. We investigate to what extent electron transfer occurs from manganese to arsenic atoms, and the strength of $d$-$d$ electron correlations. Nanoscaling and internal strain of MnAs clusters embedded in GaAs are revealed to significantly and advantageously alter the electronic structure of MnAs, triggering $d$-$d$ electron correlations, reducing the $p$-$d$ hybridization, and leading to enhanced local magnetic moments and enhanced stability of ferromagnetism, as compared to bulk MnAs.

## II - PROPERTIES OF BULK MnAs

Bulk MnAs exists in three different phases depending on temperature. Below ~313 K, MnAs crystallizes in the hexagonal NiAs-type structure.[15] This phase, α-MnAs, is ferromagnetic and shows metallic conductivity.[9,16] At ~313 K, α-MnAs transforms to orthorhombic (quasi-hexagonal) MnP-type β-MnAs, by a first-order magneto-structural phase transition.[15,17] The lattice spacing in the basal plane ($a$-axes) shrinks abruptly by 0.9% while that along the $c$-axis remains unchanged, volume being abruptly reduced by 1.8% (Ref. [15]). The Mn-As nearest-neighbor distance changes from 2.58 Å in the α phase right below the transition, to 2.52-2.61 Å in the β phase right above the transition.[17] Across the α-β phase transition, the ferromagnetic order breaks down and the resistivity jumps abruptly.[9,16] The character of the magnetic state of β-MnAs is not yet fully understood. With increasing temperature, the orthorhombic "distortion" of β-MnAs decreases continuously until reversion to the hexagonal NiAs-type structure of paramagnetic γ-MnAs at 398 K, by a second-order phase transition.[16,17] At 4.2 K, the total magnetic moment of MnAs was measured to take a value of 3.45 $\mu_B$ per formula unit,[16,18] and at room temperature it was found to be reduced to 2.85 $\mu_B$ per formula unit, of which −0.23 $\mu_B$ per formula unit correspond to magnetic polarization induced in the As sublattice.[9,19] The Mn-As nearest-neighbor distance in bulk MnAs at room temperature is 2.577 Å, while the Mn-Mn nearest-neighbor distance is 2.851 Å along the *c*-axis, and 3.719 Å in the basal planes.[20] For comparison, we note that NiAs-type MnAs nanoclusters embedded in GaAs at room temperature were found[21] to exhibit tensile strain along the *c*-axis and compressive strain in the basal planes, with the Mn-Mn distance increased up to 2.908 Å along the *c*-axis and decreased down to 3.708 Å in the basal planes, the Mn-As distance being 2.588 Å, i.e., higher than in bulk MnAs.

The valence band of MnAs is derived from the outer-shell $4s^2$ and $4p^3$ electrons of the As atoms and the $4s^2$ and $3d^5$ electrons of the Mn atoms. The 12 valence electrons are four electrons more than the eight required to form tetrahedral bonding, as in the zinc-blende structure of GaAs; therefore, to accommodate them, octahedral bonding is preferred. In the hexagonal structure of MnAs, Mn atoms are coordinated by a trigonally distorted octahedron of six As atoms. The 3*d* electrons in a Mn atom are exposed to the ligand field, with $D_{3d}$ symmetry, created by the surrounding As atoms. The cubic component of the ligand field splits the orbitally fivefold-degenerate 3*d* level in two: a less stable level for the two *e* orbitals directed toward the nearest-neighbor arsenic atoms, and a more stable level for the remaining three $t_2$ orbitals.[22] The $t_2$ level is split by the positive trigonal component of the ligand field into a more stable level for the $a_1^T$ orbital directed toward nearest-neighbor manganese atoms along the *c* axis, and a less stable level for the two $e^T$ orbitals primarily directed toward nearest-neighbor Mn atoms in the basal plane.[22] The short Mn-As distance in MnAs correlates with strong Mn-As bonding interactions. If the Mn-As bond were pure ionic, three electrons would be transferred from each Mn atom to a neighbor As atom, in order to completely fill the 4*p* shell of the electronegative As atoms. This would lead to an atomic configuration of Mn with four *d* electrons and +3 charge on Mn sites. The transfer of three electrons from Mn to As atoms in fact is not complete, the Mn-As bond actually having mixed covalent-ionic character. Mn atoms tend to retain five *d* electrons, and valence band carriers incompletely screen the positive charge on Mn sites. Hence, in the band models of MnAs, Mn atoms have a non-integer number of *d* electrons, in between 5 and 6.

Band-structure calculations, based on density-functional theory, have been carried out for the room-temperature experimental lattice parameters of bulk MnAs.[23-25] Filled As 4*s* states, hardly affected by the magnetic exchange interaction, lie at high binding energies (about –11 eV).[23-25] Hybrid Mn 3*d*-As 4*p* bands extend around the Fermi energy ($E_F$ = 0 eV), from about –6 eV up to about +4 eV.[23-25] Empty Mn 4*s* states lie above the hybrid Mn 3*d*-As 4*p* bands. Because of the magnetic exchange interaction, the spin-down *p-d* bands are shifted up in energy (non rigidly) relative to the spin-up *p-d* bands. There are little-dispersive non-bonding bands with dominant *d* orbital character and $E^T$ symmetry at about –2 eV and +1 eV for the spin-up and spin-down channels, respectively.[23,25] Bonding (antibonding) hybrid *p-d* bands extend below (above) these non-bonding states. The states crossing the Fermi energy have hybrid *p-d* character with higher *p* than *d* contribution for the spin-up channel, and nearly pure *d* character for the spin-down channel.[25-27] Due to the strong Mn 3*d*-As 4*p* covalent interactions, the As 4*p* orbitals mediate the exchange interaction between the Mn local moments.[28] Inter-atomic distances in MnAs seem to be close to critical values, across which relevant exchange interactions possibly change sign. Hence, the properties of MnAs reveal themselves extremely sensitive to small changes in the crystal structure.

## III - EXPERIMENTAL DETAILS

### A. Sample fabrication

MnAs and $Ga_{0.916}Mn_{0.084}As$ films, 120 and 117 nm-thick, respectively, were grown on heavily *n*-type Si-doped GaAs (001) substrates by molecular-beam epitaxy. MnAs growth was initiated on a carefully prepared As-rich surface of a GaAs buffer layer, grown at conventional high temperatures (550–600 °C). MnAs growth proceeded at a substrate temperature of 230 ºC, with a growth rate of 19 nm/h, and an $As_4$-to-Mn beam-equivalent-pressure ratio of ~49. The MnAs film hence grows with so-called *A* epitaxial orientation, i.e., MnAs ($1\underline{1}00$) parallel to GaAs (001), and MnAs [0001] parallel to GaAs [$1\underline{1}0$]. After MnAs growth, the sample was annealed at 286 °C for 12 min. $Ga_{0.916}Mn_{0.084}As$ growth was consecutive to the growth of a 51 nm-thick layer of LT-GaAs (GaAs grown at low temperature), on top of a *n*-type Si-doped GaAs buffer layer grown at conventional high temperature. $Ga_{0.916}Mn_{0.084}As$ growth proceeded (as LT-GaAs) at a substrate temperature of 247 ºC, with a growth rate of 5 nm/min, and an $As_4$-to-Ga beam-equivalent-pressure ratio of ~49. After $Ga_{0.916}Mn_{0.084}As$ growth, the sample was annealed first at 247 °C for 5 min, and afterwards at 640 °C for 10 min. The second annealing step at high temperature leads to the phase separation of $Ga_{0.916}Mn_{0.084}As$ into hexagonal MnAs nanocrystals embedded in an essentially Mn-depleted GaAs matrix (GaAs:MnAs). The MnAs nanocrystals have their (0001) planes oriented parallel to {111} GaAs planes, and their <$\underline{2}110$> directions oriented parallel to <110> or <$1\underline{1}0$> GaAs directions. After growth, the samples were cooled down well below room temperature and were exposed to $As_4$ flux in order to deposit protective arsenic coatings. Prior to photoemission analysis, the samples were stored under vacuum.

**B. Photoelectron emission analysis**

Photoelectron emission spectroscopic analyses were carried out on the soft-X-ray undulator beamline UE56/2-PGM-2 of the synchrotron-radiation facility BESSY II, equipped with a plane-grating monochromator. The arsenic-coated samples were briefly exposed to air, during the time necessary to cut them into pieces and to transfer them from the growth to the analysis chamber. The samples were mounted together on a variable-temperature holder, in electrical contact with a gold sheet used for calibration purposes. The arsenic coatings were desorbed by heating the samples up to ~370 °C. Photoelectron emission spectra were recorded at room temperature. Synchrotron light was incident 30º off normal, and photoemitted electrons were collected in the direction normal to the sample surface, in an angle-integrated mode, using an ESCALAB MkII electron energy analyzer. Surface cleanliness was checked by inspecting the O 1*s* and C 1*s* photoemission signals. After removal of the arsenic coatings, these signals were found to be below the detectability limit, indicating the absence of oxide and hydrocarbon contamination.

**C. Magnetometry**

After photoemission analysis, the magnetic properties of the samples were analyzed by superconducting quantum interference device (SQUID) magnetometry (Quantum Design MPMS XL). Figure 1 shows zero-field-cooled (ZFC) and field-cooled (FC) curves of the magnetization as a function of temperature in low applied magnetic fields. To measure these curves, the samples were demagnetized at 395 K and then cooled down to 5 K under zero applied magnetic field. Once at 5 K, a low magnetic field was applied along the [110] GaAs in-plane direction, with a value of 30 Oe for the sample with the continuous MnAs film, and of 50 Oe for the sample with the granular GaAs:MnAs top layer. The ZFC curve was first measured upon heating up to 395 K, and the FC curve was then measured upon cooling down to low temperature. Figure 1 shows that the Curie temperature of the magnetic MnAs nanoparticles in the granular GaAs:MnAs layer (~377 K) is remarkably higher than that for the MnAs film (~313 K). The weak thermal hysteresis observed for the MnAs film correlates with different proportions of the α and β phases upon heating and upon cooling. On the other hand, the ZFC-FC curves for the granular GaAs:MnAs layer reveal the characteristic signatures[29] of hexagonal MnAs nanocrystals embedded in a zincblende GaAs matrix: a plateau in the FC curve at low temperatures, and an abrupt decrease of the magnetization approaching the Curie temperature. Besides, the ZFC curve reveals a distribution of blocking temperatures, centered at a relatively high temperature, ~317 K, corresponding to relatively large clusters, with relatively high magnetic anisotropy, as it is characteristic of hexagonal-MnAs clusters.

## IV - Mn 2$p$ PHOTOELECTRON EMISSION SPECTRA

### A. Experimental spectra

Mn 2$p$ PES spectra for the MnAs film and for the MnAs nanoclusters embedded in GaAs are shown in Figs. 2(a) and 2(b), respectively. Binding energies are referenced to the Fermi level, which was determined from an Au 4$f$ spectrum. Under the 750-eV light excitation used, Mn 2$p$ photoelectron emission appears superimposed on backgrounds of secondary-electron emission (dashed curves) that we determined by smoothly extrapolating the trend revealed in wide spectra (not shown) within the Mn 2$p$ spectral region. Figure 2(c) compares the two Mn 2$p$ spectra, after subtraction of the respective secondary-electron backgrounds and area normalization. The spectrum of the continuous film essentially consists of a 2$p_{3/2}$-2$p_{1/2}$ spin-orbit split doublet. The individual lines are observed to have marked tails on their high binding-energy sides, as previously observed for Mn-metal and for Mn-containing compounds.[30,31] The line asymmetry arises from intra-atomic multiplet effects,[32] and core-hole screening processes,[33] similarly to what has been shown for Mn-metal.[31]

The Mn $2p$ spectrum of the nanoclusters is more complex: pronounced satellite structures ["s" in Fig. 2(c)] show up at 3-5 eV higher binding energies than the main peaks ("m") of the spin-orbit doublet. These satellite structures do not correspond to chemically shifted oxide components, as guaranteed by the absence of O $1s$ signal; rather they are satellites induced by electron correlation effects. The main (m) and satellite (s) emissions correspond to different photoemission final states, with different $d$-orbital occupation, possible because of the incomplete screening of the Mn $2p$ core hole by Mn $3d$-As $4p$ valence electrons. Whereas the absence of "s" satellite structures for the film indicates efficient core-hole screening, the presence of satellite structures for the nanoclusters reveals poor core-hole screening and relatively strong $d$-$d$ electron correlations.

Energy losses ["p" in Fig. 2(c)] due to plasmon excitation are observed to contribute to both film and nanocluster spectra, at roughly 19 eV higher binding energies than the main peaks (m) of the photoemission doublet. These (intrinsic) plasmons are excited in the process of screening of the Mn $2p$ core hole. Plasmon losses are seen to be somewhat more intense for the nanoclusters than for the film.

**B. Method of simulation**

In order to gain insight into the electronic correlations in the MnAs nanoclusters, we analyze their Mn $2p$ photoemission lineshape on the basis of a CI cluster model.[34] In the cluster approximation, we consider an idealized structure, a MnAs$_6$ cluster, with a central Mn cation surrounded by six As anions, arranged in a NiAs-type MnAs structure with the lattice parameters previously measured[21] for hexagonal MnAs nanoclusters embedded in GaAs. We concentrate on energy levels arising from hybridization of Mn $3d$ and As $4p$ orbitals. We assume that the pure ionic configuration of the Mn cation is the high-spin $d^4$ configuration, i.e., the $Mn^{3+}$ state with local moment $S=2$. The wave functions for the *initial* states (states before Mn 2p photoemission) are described in the CI picture as linear combinations of the pure ionic configuration and screened charge-transfer configurations, in which one or more electrons are transferred to the Mn 3d levels from ligand (As) orbitals:

$$\Psi_{i,k} = a_{k,0}\left|d^4\right\rangle + \sum_m a_{k,m}\left|d^{4+m}\underline{L}^m\right\rangle, \tag{1}$$

with

$$\left|a_{k,0}\right|^2 + \sum_m \left|a_{k,m}\right|^2 = 1. \tag{2}$$

Here, $\underline{L}$ stands for a hole in a ligand (As) orbital and the summation stands for all the charge-transfer configurations $\left|d^{4+m}\underline{L}^m\right\rangle$ with the same symmetry as the ionic configuration $\left|d^4\right\rangle$. The number of 3d electrons on the Mn atom is given by:

$$N^e_{Mn3d}(\Psi_{i,k}) = 4\left|a_{k,0}\right|^2 + \sum_m (4+m)\left|a_{k,m}\right|^2, \tag{3}$$

and the number of 4p holes on the As atom is given by:

$$N^h_{As4p}(\Psi_{i,k}) = \sum_m m\left|a_{k,m}\right|^2. \tag{4}$$

The wave functions for the *final* states (states after Mn 2p photoemission) are given by:

$$\Psi_{f,k} = b_{k,0}\left|\underline{2p}\,d^4\right\rangle + \sum_m b_{k,m}\left|\underline{2p}\,d^{4+m}\underline{L}^m\right\rangle, \tag{5}$$

with

$$\left|b_{k,0}\right|^2 + \sum_m \left|b_{k,m}\right|^2 = 1. \tag{6}$$

Here, $\underline{2p}$ stands for a Mn 2p core hole. We account for crystal-field effects and for the *d-d* exchange interaction by extending the basis set to distinguish between each individual orbital symmetry and spin state: $a_1^T \uparrow, a_1^T \downarrow, e^T \uparrow, e^T \downarrow, e \uparrow,$ and $e \downarrow$. The wave functions for the initial states are then:

$$\Psi_{i,k} = a_{k,0} \left| (\underline{e} \downarrow)^1 (\underline{a}_1^T \uparrow)^1 (\underline{e}^T \uparrow)^2 (\underline{e} \uparrow)^2 \right\rangle$$
$$+ \sum_{m_3',m_1',m_2',m_3} a_{k,m_3 m_1' m_2' m_3'} \left| (\underline{e} \downarrow)^{1-m_3} (\underline{L}_e \downarrow)^{m_3} (\underline{a}_1^T \uparrow)^{1-m_1'} (\underline{L}_{a_1^T} \uparrow)^{m_1'} (\underline{e}^T \uparrow)^{2-m_2'} (\underline{L}_{e^T} \uparrow)^{m_2'} (\underline{e} \uparrow)^{2-m_3'} (\underline{L}_e \uparrow)^{m_3'} \right\rangle \quad (7)$$

Here, $\underline{a}_1^T, \underline{e}^T,$ and $\underline{e}$ denote holes on Mn $3d$ orbitals with $A_1^T, E^T,$ and $E$ symmetry, respectively, and $\underline{L}_{a_1^T}, \underline{L}_{e^T},$ and $\underline{L}_e$ denote holes on ligand orbitals with $A_1^T, E^T,$ and $E$ symmetry, respectively. The $L_{a_1^T}, L_{e^T},$ and $L_e$ ligand orbitals are combinations of $4p$ orbitals on the 6 neighbor As atoms. $m_1'$, $m_2'$, $m_3'$, and $m_3$ indicate the number of $\underline{L}_{a_1^T} \uparrow, \underline{L}_{e^T} \uparrow, \underline{L}_e \uparrow,$ and $\underline{L}_e \downarrow$ holes, respectively. The right-hand side is summed over all possible combinations of symmetry-matching ligand holes. The 36 basis functions involved are given by Slater determinants. The 13 first basis functions, up to 2 ligand holes, are listed in Table I.

The diagonal matrix elements of the *initial-state* Hamiltonian, $H^i$, are given by (Ref. [34]):

$$\begin{aligned}
\langle d^4 | H^i | d^4 \rangle &= 0 \quad \text{(reference)} \\
\langle d^5 \underline{L} | H^i | d^5 \underline{L} \rangle &= \varepsilon(d^5 \underline{L}) - \varepsilon(d^4) + \Delta \\
\langle d^6 \underline{L}^2 | H^i | d^6 \underline{L}^2 \rangle &= \varepsilon(d^6 \underline{L}^2) - \varepsilon(d^4) + 2\Delta + U \\
&\vdots \\
\langle d^{4+m} \underline{L}^m | H^i | d^{4+m} \underline{L}^m \rangle &= \varepsilon(d^{4+m} \underline{L}^m) - \varepsilon(d^4) + m\Delta + \tfrac{1}{2} m(m-1) U \\
&\vdots \\
\langle d^{10} \underline{L}^6 | H^i | d^{10} \underline{L}^6 \rangle &= \varepsilon(d^{10} \underline{L}^6) - \varepsilon(d^4) + 6\Delta + 15U
\end{aligned} \quad (8)$$

where $\varepsilon(d^{4+m} \underline{L}^m)$ is the Coulomb-exchange interaction energy of the configuration $d^{4+m}$, $\Delta$ is the energy required to transfer one electron from a ligand orbital to a Mn $3d$ orbital, $\Delta \equiv E(d^{n+1} \underline{L}) - E(d^n)$, and $U$ is the on-site $d$-$d$ Coulomb interaction energy, $U \equiv E(d^{n-1}) + E(d^{n+1}) - 2E(d^n)$. $\Delta$ and $U$ are defined with respect to the center of gravity of the $d^n$ multiplet. The $d$-$d$ Coulomb-exchange interaction energy for a configuration is expressed[34] in terms of the Kanamori parameters[35] $u'$, $u$ and $j$, and these in turn are expressed[34] in terms of the Racah parameters[36] $A$, $B$, and $C$ ($A \equiv U + \frac{14}{9}B - \frac{7}{9}C$). As previously done,[34] we fix $B$ and $C$ at the values corresponding to the free Mn ion $B$=0.120 eV, $C$=0.552 eV (Ref. [37]). The 13 first diagonal matrix elements of the initial-state Hamiltonian are listed in Table I.

Hybridization between the Mn 3$d$ and As 4$p$ orbitals is treated, in the CI picture, as a perturbation.[34] It is included via one-electron mixing (off-diagonal) matrix elements, via transfer integrals. To account for the anisotropic $p$-$d$ hybridization in MnAs, we consider different transfer integrals for the orbitals with $A_1^T$, $E^T$, and $E$ symmetries:

$$T_{a_1^T} \equiv \langle a_1^T | H^i | L_{a_1^T} \rangle$$
$$T_{e^T} \equiv \langle e^T | H^i | L_{e^T} \rangle \quad (9)$$
$$T_e \equiv \langle e | H^i | L_e \rangle$$

For bulk hexagonal MnAs, whose lattice parameters are $a$=3.719 Å and $c$=5.702 Å (Ref. [20]), the transfer integrals are expressed in terms of the Slater-Koster parameters $(pd\sigma)$ and $(pd\pi)$ as:

$$T_{a_1^T} = 0.01(pd\sigma) - 1.94(pd\pi)$$
$$T_{e^T} = 0.84(pd\sigma) - 0.98(pd\pi) \quad (10)$$
$$T_e = 1.29(pd\sigma) - 0.43(pd\pi)$$

and for hexagonal MnAs nanocrystals embedded in GaAs, with slightly different lattice parameters $a$=3.708 Å and $c$=5.816 Å because of their strained state,[21] the transfer integrals are expressed as:

$$T_{a_1^T} = 0.005(pd\sigma) - 1.97(pd\pi)$$
$$T_{e^T} = 0.82(pd\sigma) - 0.98(pd\pi) \quad (11)$$
$$T_e = 1.30(pd\sigma) - 0.42(pd\pi)$$

The transfer integrals are seen to be only slightly sensitive to the typical structural distortion of hexagonal MnAs nanocrystals embedded in GaAs. Assuming the relationship $(pd\sigma)/(pd\pi) = -2.16$, as previously done,[14,38] the transfer integrals for the strained nanoclusters are found to be given by: $T_{a_1^T} = 0.92(pd\sigma)$, $T_{e^T} = 1.28(pd\sigma)$, and $T_e = 1.49(pd\sigma)$. Note that p-d hybridization is highest for orbitals with $E$ symmetry, consistent with these orbitals being mostly directed towards As anions. Table I lists the non-zero off-diagonal elements of the 13×13 leading submatrix of the initial-state Hamiltonian, in terms of the transfer integrals.

The photoemission final states differ from the initial states by the Coulomb attraction $Q$ between the Mn 2p core hole and the Mn 3d-Mn 4p valence electrons, which pulls down the charge-transferred states relative to the pure ionic state. With the presence of a Mn 2p core hole, the diagonal matrix elements of the *final-state* Hamiltonian $H^f$ are given by (Ref. [34]):

$$\begin{aligned}
\langle \underline{2p}d^4 | H^f | \underline{2p}d^4 \rangle &= E_c \\
\langle \underline{2p}d^5\underline{L} | H^f | \underline{2p}d^5\underline{L} \rangle &= E_c + \varepsilon(d^5\underline{L}) - \varepsilon(d^4) + \Delta - Q \\
\langle \underline{2p}d^6\underline{L}^2 | H^f | \underline{2p}d^6\underline{L}^2 \rangle &= E_c + \varepsilon(d^6\underline{L}^2) - \varepsilon(d^4) + 2(\Delta - Q) + U \\
&\vdots \\
\langle \underline{2p}d^{4+m}\underline{L}^m | H^f | \underline{2p}d^{4+m}\underline{L}^m \rangle &= E_c + \varepsilon(d^{4+m}\underline{L}^m) - \varepsilon(d^4) + m(\Delta - Q) + \tfrac{1}{2}m(m-1)U \\
&\vdots \\
\langle \underline{2p}d^{10}\underline{L}^6 | H^f | \underline{2p}d^{10}\underline{L}^6 \rangle &= E_c + \varepsilon(d^{10}\underline{L}^6) - \varepsilon(d^4) + 6(\Delta - Q) + 15U
\end{aligned} \quad (12)$$

where $E_c$ is the energy of the Mn 2p core-hole ($|\underline{2p}d^4\rangle$) relative to the ionic state ($|d^4\rangle$). As previously done,[14,34] we fix the average attractive Coulomb energy at $Q=1.25U$. We assume that the off-diagonal matrix elements for the final-state Hamiltonian are the same as those for the initial-state Hamiltonian.

After finding the eigenfunctions ($\tilde{\Psi}_{i,l}$ and $\tilde{\Psi}_{f,k}$) and eigenvalues ($\tilde{E}_{i,l}$ and $\tilde{E}_{f,k}$) for the initial and final states, diagonalizing the respective Hamiltonians, the photoemission spectrum is given, in the sudden approximation, by (Ref. [34]):

$$\rho(e) = \sum_{k} \left| \langle \tilde{\Psi}_{f,k} | \underline{2p} | \tilde{\Psi}_{i,g} \rangle \right|^2 \delta(h\nu - E_{kin} - \tilde{E}_{f,k}), \tag{13}$$

summed over all final states. Here, $\underline{2p}$ is the annihilation operator of a Mn 2p core electron in the initial state, $\tilde{\Psi}_{f,k}$ is the $k$ final-state eigenfunction, $\tilde{\Psi}_{i,g}$ is the *ground* initial-state eigenfunction, $\tilde{E}_{f,k}$ is the $k$ final-state eigenenergy, $h\nu$ is the photon energy, and $E_{kin}$ is the kinetic energy of the photoelectron. Hence, the spectral weight of each final-state eigenenergy $\tilde{E}_{f,k}$ is given by the square of the overlap integral between $|\tilde{\Psi}_{f,k}\rangle$ and $|\underline{2p}\tilde{\Psi}_{i,g}\rangle$, where $|\underline{2p}\tilde{\Psi}_{i,g}\rangle$ represents the frozen state obtained by annihilating a Mn 2p core electron in the ground initial state. To simulate the experimental spectrum, the theoretical elastic-photoemission δ-lines are broadened by functions representing the experimental resolution and intrinsic photoemission lineshape. We represent elastic-photoemission lines by Gaussian/Doniach-Sunjic convoluted functions, which account for the core-hole lifetime, electron-hole pair excitation, multiplet splitting, and experimental resolution. We fix the core-level asymmetry at the value $\alpha$=0.41 eV, previously considered for Mn compounds.[30] The parameters $U$, $\Delta$, and $pd\sigma$ are varied to provide a best fit to the experimental data.

We represent quantized (intrinsic) plasmon energy losses by Voigt doublet profiles.[39] On the other hand, we simulate the contribution of inelastically scattered photoelectrons, both zero-energy-loss photoelectrons and those which have previously experienced quantized plasmon energy losses, by Tougaard-type integral backgrounds.[40] For the inelastic-scattering cross section, we use the coefficients $\tilde{C}$=1000 eV$^2$ and $\tilde{D}$=13300 eV$^2$, corresponding to the universal cross section for transition metals,[40] which we have seen describe well the background in any spectral region far from the secondary-electron region.[41] To better account for unquantized (extrinsic) plasmon energy losses, we leave the inelastic-scattering $\widetilde{B}$ coefficient as a fitting parameter.[42] We simulated the spectral lineshapes with the aid of the CasaXPS software package.

### C. Spectral components

Figure 3(b) shows the analysis of the Mn 2$p$ photoemission lineshape for the MnAs nanoclusters, based on the CI cluster model. The $\Delta$, $U$, and $pd\sigma$ best-fit values are listed in Table II. With these values, the CI cluster model most well reproduces the experimental spectrum, including the satellite structures. The leading Mn $2p_{3/2}$ photoemission signal ("m") peaks at 641.15 eV [Fig. 2(c)]. Best fitting is achieved when the spin-orbit splitting is set to 11.4 eV, the elastic-photoemission Gaussian width to 3.6 eV, and the inelastic-scattering $\widetilde{B}$ coefficient of the Tougaard background to 1800 eV$^2$. The CI analysis reveals that the leading peaks of the Mn 2$p$ elastic photoemission for the nanoclusters [spin-orbit split "m" lines in Fig. 3(b)] correspond to transitions to a final state mixing the following main configurations:

$$
\begin{aligned}
22\% \ &|\underline{2p}d^6L^2\rangle = |\underline{2p}(a_1^T\uparrow)^1(e^T\uparrow)^2(e\uparrow)^2(e\downarrow)^1(\underline{L}_e\downarrow)^1(\underline{L}_e\uparrow)^1\rangle \\
19\% \ &|\underline{2p}d^6L^2\rangle = |\underline{2p}(a_1^T\uparrow)^1(e^T\uparrow)^2(e\uparrow)^2(e^T\downarrow)^1(\underline{L}_e\downarrow)^1(\underline{L}_{e^T}\uparrow)^1\rangle \qquad (14)\\
18\% \ &|\underline{2p}d^5L\rangle \ = |\underline{2p}(a_1^T\uparrow)^1(e^T\uparrow)^2(e\uparrow)^2(\underline{L}_e\downarrow)^1\rangle
\end{aligned}
$$

and additional small contributions from other configurations. On the other hand, the satellite structures ["s" in Fig. 2(c)] are revealed to consist of several contributions, the most intense ones being "s1" and "s2", at 5.2 and 3.4 eV higher binding energies than the leading "m" lines, respectively [Fig. 3(b)]. The "s1" lines correspond to transitions to a final state mixing the following main configuration:

$$
34\% \ |\underline{2p}d^5L\rangle = |\underline{2p}(a_1^T\uparrow)^1(e^T\uparrow)^2(e\uparrow)^2(\underline{L}_e\downarrow)^1\rangle \qquad (15)
$$

and additional small contributions from other configurations. The "s2" lines correspond to transitions to a final state mixing the following main configurations:

$$
\begin{aligned}
19\% \quad &\left|\underline{2p}\,d^5\,L\right\rangle = \left|\underline{2p}(a_1^T\uparrow)^1(e^T\uparrow)^2(e\uparrow)^2(L_e\downarrow)^1\right\rangle \\
15\% \quad &\left|\underline{2p}\,d^6\,\underline{L}^2\right\rangle = \left|\underline{2p}(a_1^T\uparrow)^1(e^T\uparrow)^2(e\uparrow)^1(e^T\downarrow)^1(e\downarrow)^1(\underline{L}_{e^T}\uparrow)^1(\underline{L}_e\uparrow)^1\right\rangle
\end{aligned}
\qquad (16)
$$

and additional small contributions from other configurations. In addition to elastic photoemission, a doublet signal ($p_b$), assumed to correspond to the excitation of a *bulk* plasmon, with a characteristic Gaussian width of 4.3 eV, is found to contribute to the Mn 2$p$ nanocluster spectrum at 18.75 eV higher binding energy than the main doublet. Besides, a less intense doublet signal ($p_i$), assumed to correspond to the excitation of an *interface* plasmon, with a characteristic Gaussian width of 3.6 eV, is found to contribute at 14 eV higher binding energy than the main doublet. The distinguishable additional contribution of an interface plasmon loss is understandable, considering the large number of Mn atoms located in MnAs/GaAs interface environments in the nanocluster morphology.

The Mn 2$p$ spectrum of the continuous MnAs film [Fig. 3(a)] can be reasonably well described considering the contributions of (i) a leading spin-orbit split doublet, accounting for elastic photoemission, (ii) a small $p_b$ doublet accounting for a quantized plasmon energy loss, and (iii) a Tougaard-type background. The leading Mn 2$p_{3/2}$ photoemission signal peaks at 640.05 eV [Fig. 2(c)]. Best fitting is achieved when the spin-orbit splitting is set to 11.25 eV, the elastic-photoemission Gaussian width to 2.7 eV, and the inelastic-scattering $\tilde{B}$ coefficient to 3250 eV$^2$. The $p_b$ doublet signal, assumed to correspond to the excitation of a bulk plasmon, with Gaussian width of 6.75 eV, is found to be centered at 19.4 eV higher binding energy than the main doublet. This plasmon energy loss is to be compared with the screened plasma frequency calculated by Ravindran et al. (22.2 eV) and that derived by Bärner et al. (17.0 eV) from reflectivity spectra.[43,44]

# V - MnAs ELECTRONIC AND MAGNETIC PROPERTIES

The energy-level diagram for the initial and final states of MnAs nanoclusters embedded in GaAs, derived from the CI analysis, is shown in Fig. 4. If *p-d* hybridization (covalency of the Mn-As bond) were disregarded, i.e., if the Mn-As bond were considered to be 100% ionic, the $|d^5 \underline{L}\rangle$ configuration would be found to be the ground initial state (left-side diagram in Fig. 4), consistent with the negative value of $\Delta$. However, as a consequence of the hybrid ionic-covalent nature of the MnAs bonding interactions, of the effective *p-d* hybridization, the ground state is a mixture of charge-transfer configurations. From the CI analysis we obtain the following mixture of configurations for the ground initial state:

$$
\begin{aligned}
60\% \quad &|d^5 \underline{L}\rangle = |(a_1^T \uparrow)^1 (e^T \uparrow)^2 (e \uparrow)^2 (\underline{L}_e \downarrow)^1\rangle \\
12\% \quad &|d^6 \underline{L}^2\rangle = |(a_1^T \uparrow)^1 (e^T \uparrow)^2 (e \uparrow)^2 (e \downarrow)^1 (\underline{L}_e \downarrow)^1 (\underline{L}_e \uparrow)^1\rangle \\
9\% \quad &|d^6 \underline{L}^2\rangle = |(a_1^T \uparrow)^1 (e^T \uparrow)^2 (e \uparrow)^2 (e^T \downarrow)^1 (\underline{L}_e \downarrow)^1 (\underline{L}_{e^T} \uparrow)^1\rangle \\
7\% \quad &|d^4\rangle = |(a_1^T \uparrow)^1 (e^T \uparrow)^2 (e \uparrow)^1\rangle \\
3\% \quad &|d^5 \underline{L}\rangle = |(a_1^T \uparrow)^1 (e^T \uparrow)^2 (e \uparrow)^1 (e \downarrow)^1 (\underline{L}_e \uparrow)^1\rangle \\
2\% \quad &|d^5 \underline{L}\rangle = |(a_1^T \uparrow)^1 (e^T \uparrow)^2 (e \uparrow)^1 (e^T \downarrow)^1 (\underline{L}_{e^T} \uparrow)^1\rangle \\
2\% \quad &|d^6 \underline{L}^2\rangle = |(a_1^T \uparrow)^1 (e^T \uparrow)^2 (e \uparrow)^2 (a_1^T \downarrow)^1 (\underline{L}_e \downarrow)^1 (\underline{L}_{a_1^T} \uparrow)^1\rangle
\end{aligned}
\quad (17)
$$

and additional very small contributions from other configurations. The ground initial state of the MnAs nanoclusters is found to be largely dominated by the $|d^5 \underline{L}\rangle = |(a_1^T \uparrow)^1 (e^T \uparrow)^2 (e \uparrow)^2 (\underline{L}_e \downarrow)^1\rangle$ configuration, which maximizes the local Mn moment and involves a hole in a spin-up valence orbital with *E* symmetry at the $\Gamma$ point. The number of *d* electrons on the Mn atom is found to be $N^e_{\text{Mn}\,3d}$=5.22, and the number of *p* holes on the As atom to be $N^h_{\text{As}\,4p}$=1.22. Hence, Mn cations are found not to have 3+ valence and 4 *d* electrons, as assumed by the ionic models, but +1.78 effective valence and 5.22 *d* electrons. The mixture of configurations for the ground initial state indicates that the spin-up *e* orbital acts as a channel maximizing the local Mn moment, and the spin-down *e* and $e^T$ orbitals act as channels screening the local Mn charge, with the concomitant effect of reducing the local Mn moment.

Insight into the effective charge of Mn cations in MnAs can also be gained from the measured Mn $2p$ binding energies. When atoms in compounds have a high ionic charge and a localized electronic structure, the corresponding core-level lines exhibit large chemical shifts in photoemission spectra with respect to the elemental (metallic) material. On the contrary, when atoms participate in extended bonding, the core-level lines exhibit low chemical shifts, since delocalization of valence electrons results in greater nuclear screening (lower effective charge). If cation- and anion-based orbitals strongly hybridize because of a covalent nature of the bonding interactions, the binding energy of the cation atoms in the compound approaches that in the elemental material. Since the Mn $2p_{3/2}$ binding energies measured for the MnAs film (640.05 eV) and for the MnAs nanoclusters embedded in GaAs (641.15 eV) are higher than that (638.7-638.9 eV) reported for Mn-metal,[31,45] we conclude that Mn atoms in MnAs films and nanoclusters are definitely in a cationic state, i.e., there are effective electron transfers from the Mn cations to the As anions (1.78 electrons per Mn atom for the nanoclusters as derived from the CI analysis). The larger Mn $2p_{3/2}$ chemical shift measured for the nanoclusters evidences the higher ionicity (lower covalency) of the Mn-As bonding interactions (lower $p$-$d$ hybridization), as compared to the film (to bulk MnAs). Analysis of the measured Mn $2p$ spin-orbit splittings leads to the same conclusions on the effective Mn-cation charge as those derived from analysis of the measured Mn $2p_{3/2}$ binding energies. The spin-orbit splitting is inversely proportional to the ionic radius,[46] the latter decreasing with increasing positive charge in the cations. The higher Mn $2p$ spin-orbit splitting measured for the nanoclusters (11.4 eV) as compared to that measured for the film (11.25 eV) indicates higher effective positive charge of the Mn cations for the nanoclusters than that for the film. On the basis of the above discussion, we expect the number of $d$ electrons on the Mn atom for bulk MnAs to be somewhat higher than that for the MnAs nanoclusters, $N^e_{\text{Mn}\,3d}$ =5.22, and somewhat lower than that reported[31] for Mn-metal, $N^e_{\text{Mn}\,3d}$ =5.7. Accordingly, we expect the effective charge of Mn cations in bulk MnAs to be somewhat lower than +1.78.

The presence of correlation satellites in the Mn 2p spectrum of the MnAs nanoclusters, and their weakness or absence in the spectrum of the MnAs film, indicates that on-site d-d Coulomb interactions are relatively strong in the nanoclusters, and weak in the film. The enhancement of the d-d Coulomb interactions in the nanoclusters is assumed to correlate with increased localization of the Mn 3d orbitals, i.e., with a reduced Mn 3d bandwidth, as compared to bulk MnAs material. Because of the increased d-electron localization, Mn 3d charge fluctuations in the nanoclusters are slowed down to such an extent that several photoemission final states, with different occupation of the d orbitals, become apparent in photoemission spectra. The combination of two factors is assumed to be responsible for the increased localization of the Mn 3d electrons in the nanoclusters: (i) the MnAs size reduction, and (ii) the weakened p-d hybridization as a result of the internal strain.

The local magnetic moment is expected to be enhanced in the nanoclusters, as compared to the film (to bulk MnAs), as a consequence of the shrinking of the d band. The enhanced d orbital localization is expected to result in an increased exchange splitting between the majority and minority channels of the Mn d states, thus in higher occupation of the majority states and lower occupation of the minority states. An analysis of the measured plasmon-loss intensities, which scale with the number of spin-unpaired valence electrons,[42] supports this idea. The higher plasmon-loss intensity measured for the nanoclusters is consistent with an enhanced local magnetic moment.

As a consequence of the enlarged volume when going from MnAs to MnBi, the p character of the states crossing the Fermi energy was found to be enhanced and the spin-up d band was found to be lowered, leading to enhanced stability of ferromagnetism.[27,43] The enlarged unit-cell volume of MnAs nanoclusters embedded in GaAs, as compared to bulk MnAs, is expected to similarly lead to enhanced *p* character of the states crossing the Fermi energy. This idea is supported by the measured bulk plasmon loss energies. The plasmon loss energy is sensitive to the density of Mn 3*d* valence electrons: a greater density leads to a higher Mn 2*p* plasmon energy.[42] The lower bulk plasmon energy measured for the nanoclusters is consistent with a reduced (Mn) *d* character [enhanced (As) *p* character] of the states crossing the Fermi energy. Accordingly, the enlarged unit-cell volume would be responsible for the enhanced stability of ferromagnetism in embedded MnAs nanocrystals.

Contrary to the case of MnAs films epitaxially grown on GaAs substrates, the structural distortion of MnAs nanoclusters embedded in GaAs plays an advantageous role in stabilizing ferromagnetism. Whereas epitaxy of MnAs films on GaAs substrates stabilizes the non-ferromagnetic β phase below its stability range in bulk MnAs,[47] the embedded morphology of the MnAs nanoclusters stabilizes the ferromagnetic α phase above its stability range in bulk MnAs. It remains to be shown whether an octahedrally distorted β phase exists or is suppressed for MnAs nanoclusters embedded in GaAs, and how ferromagnetism in the MnAs nanoclusters destabilizes upon heating, whether as a first-order α-β phase transition, like in bulk MnAs, or as a second-order direct α-γ phase transition.

## VI - CONCLUSIONS

We have carried out a comparative photoemission spectroscopy (PES) study of the electronic structure of hexagonal MnAs, as epitaxial film on GaAs(001) and as nanoclusters embedded in GaAs, focusing on the effect of nanoscaling and internal nanocluster strain on Mn-Mn and Mn-As bonding interactions.

Mn 2$p$ photoemission reveals that on-site $d$-$d$ Coulomb interactions are relatively strong for the nanoclusters and weak for the film. The Mn 2$p$ photoelectron emission spectrum of the embedded nanoclusters shows a complex structure, including correlation satellites, that we have analyzed using a configuration-interaction (CI) cluster model. We find the number of $d$ electrons on the Mn atom for the nanoclusters to be ~5.22, and the effective Mn-cation charge to be about +1.78. The ground state of the nanoclusters is found to be largely dominated by a $\left|d^5 L\right\rangle$ configuration that maximizes the local Mn moment and involves a hole in a spin-up valence orbital with $E$ symmetry at the $\Gamma$ point.

Nanoscaling and internal strain in the nanocrystals significantly alter the properties of MnAs. The internal nanocluster strain leads to a longer Mn-As nearest-neighbor distance that results in reduced $p$-$d$ hybridization and enhanced ionic character of the Mn-As bond, as compared to bulk MnAs. The spatial confinement and weakened $p$-$d$ hybridization in the nanoclusters enhance $d$-electron localization, triggering $d$-$d$ electron correlations and enhancing the local Mn moments.

The Curie temperature of the nanoclusters is measured to be remarkably higher than that of the film and of bulk MnAs. We believe that the enhanced stability of ferromagnetism in the embedded MnAs nanoclusters correlates to the increased $p$ character of the states crossing the Fermi energy, as a result of the enlarged nanocluster unit-cell volume. Contrary to epitaxial MnAs films on GaAs, the mechanical properties of the granular GaAs:MnAs composite advantageously modify the magnetic properties of MnAs, stabilizing the ferromagnetic α phase above its stability range in bulk MnAs, well above room temperature.

## ACKNOWLEDGMENTS

We gratefully acknowledge A. Kumar and M. Tallarida for assistance to run the photoemission experiments, and the BESSY staff for support. This work has been partly supported by the Spanish Ministry of Education and Science ("Ramón y Cajal" program and MAT2007–66719 grant).

**FIGURE CAPTIONS**

**Figure 1.** Temperature dependence of the magnetization M measured under low applied field, normalized by the value of the field-cooled magnetization at $T=5$ K ($M_0$), for an epitaxial MnAs film grown on GaAs (grey curves) and for MnAs nanoclusters embedded in GaAs (black curves). The samples were first cooled from 395 K down to 5 K under zero applied field. Then, a low field was applied (30 Oe for the film, and 50 Oe for the nanoclusters) and the zero-field-cooled (zfc) and field-cooled (fc) curves were consecutively measured, upon heating and upon cooling, respectively.

**Figure 2.** (a) and (b) Mn 2$p$ raw photoemission spectra for a MnAs film (grey dots) and for MnAs nanoclusters embedded in GaAs (black dots), respectively. The secondary-electron backgrounds are shown by dashed curves. (c) The same spectra after background subtraction and area normalization. Binding energies are referenced to the Fermi level (the negative sign is omitted).

**Figure 3.** (Color online) Lineshape analysis of the Mn 2$p$ spectra shown in Fig. 2(c) for (a) MnAs film and for (b) MnAs nanoclusters embedded in GaAs. The vertical blue (black) bars are the calculated unbroadened spectral lines corresponding to elastic photoemission, either main (m) or satellite (s1, s2) contributions. The vertical green (grey) bars are unbroadened spectral lines corresponding to the excitation of bulk ($p_b$) or interface ($p_i$) plasmons. The corresponding broadened spectral contributions are shown by dotted blue (black) curves (elastic photoemission) and dash-dotted green (grey) curves (plasmon losses), superimposed on the respective background contributions. The grey curves represent the theoretical spectra obtained adding up all contributions.

**Figure 4.** Energy-level diagrams for the initial (before photoemission) and final (after Mn 2$p$ photoemission) states of the MnAs$_6$ cluster, assuming a nominal Mn$^{+3}$ oxidation state.

**TABLE I.** Partial listing (up to 2 ligand holes) of the basis functions and Hamiltonian matrix elements for the initial state of a MnAs$_6$ model cluster, assuming nominal Mn$^{3+}$ oxidation state (high-spin $d^4$ configuration).

| Basis functions | Hamiltonian matrix elements |
|---|---|
| $\phi_1 = \left\|(a_1^T \uparrow)^1 (e^T \uparrow)^2 (e \uparrow)^1\right\rangle$ | $H_{1,1} = -\frac{35}{3}B - \frac{14}{3}C$ |
| $\phi_2 = \left\|(a_1^T \uparrow)^1 (e^T \uparrow)^2 (e \uparrow)^1 (a_1^T \downarrow)^1 (\underline{L}_{a_1^T} \uparrow)^1\right\rangle$ | $H_{2,2} = -\frac{40}{9}B - \frac{16}{9}C + \Delta$ |
| $\phi_3 = \left\|(a_1^T \uparrow)^1 (e^T \uparrow)^2 (e \uparrow)^1 (e^T \downarrow)^1 (\underline{L}_{e^T} \uparrow)^1\right\rangle$ | $H_{3,3} = -\frac{40}{9}B - \frac{16}{9}C + \Delta$ |
| $\phi_4 = \left\|(a_1^T \uparrow)^1 (e^T \uparrow)^2 (e \uparrow)^2 (\underline{L}_e \downarrow)^1\right\rangle$ | $H_{4,4} = -\frac{175}{9}B - \frac{70}{9}C + \Delta$ |
| $\phi_5 = \left\|(a_1^T \uparrow)^1 (e^T \uparrow)^2 (e \uparrow)^1 (e \downarrow)^1 (\underline{L}_e \uparrow)^1\right\rangle$ | $H_{5,5} = -\frac{40}{9}B - \frac{16}{9}C + \Delta$ |
| $\phi_6 = \left\|(a_1^T \uparrow)^1 (e^T \uparrow)^2 (e \uparrow)^1 (a_1^T \downarrow)^1 (e^T \downarrow)^1 (\underline{L}_{a_1^T} \uparrow)^1 (\underline{L}_{e^T} \uparrow)^1\right\rangle$ | $H_{6,6} = +\frac{5}{6}B + \frac{1}{3}C + 2\Delta + U$ |
| $\phi_7 = \left\|(a_1^T \uparrow)^1 (e^T \uparrow)^2 (e \uparrow)^2 (a_1^T \downarrow)^1 (\underline{L}_e \downarrow)^1 (\underline{L}_{a_1^T} \uparrow)^1\right\rangle$ | $H_{7,7} = -\frac{35}{3}B - \frac{14}{3}C + 2\Delta + U$ |
| $\phi_8 = \left\|(a_1^T \uparrow)^1 (e^T \uparrow)^2 (e \uparrow)^1 (a_1^T \downarrow)^1 (e \downarrow)^1 (\underline{L}_{a_1^T} \uparrow)^1 (\underline{L}_e \uparrow)^1\right\rangle$ | $H_{8,8} = +\frac{5}{6}B + \frac{1}{3}C + 2\Delta + U$ |
| $\phi_9 = \left\|(a_1^T \uparrow)^1 (e^T \uparrow)^2 (e \uparrow)^1 (e^T \downarrow)^2 (\underline{L}_{e^T} \uparrow)^2\right\rangle$ | $H_{9,9} = +\frac{5}{6}B + \frac{1}{3}C + 2\Delta + U$ |
| $\phi_{10} = \left\|(a_1^T \uparrow)^1 (e^T \uparrow)^2 (e \uparrow)^2 (e^T \downarrow)^1 (\underline{L}_e \downarrow)^1 (\underline{L}_{e^T} \uparrow)^1\right\rangle$ | $H_{10,10} = -\frac{35}{3}B - \frac{14}{3}C + 2\Delta + U$ |
| $\phi_{11} = \left\|(a_1^T \uparrow)^1 (e^T \uparrow)^2 (e \uparrow)^1 (e^T \downarrow)^1 (e \downarrow)^1 (\underline{L}_{e^T} \uparrow)^1 (\underline{L}_e \uparrow)^1\right\rangle$ | $H_{11,11} = +\frac{5}{6}B + \frac{1}{3}C + 2\Delta + U$ |
| $\phi_{12} = \left\|(a_1^T \uparrow)^1 (e^T \uparrow)^2 (e \uparrow)^2 (e \downarrow)^1 (\underline{L}_e \downarrow)^1 (\underline{L}_e \uparrow)^1\right\rangle$ | $H_{12,12} = -\frac{35}{3}B - \frac{14}{3}C + 2\Delta + U$ |
| $\phi_{13} = \left\|(a_1^T \uparrow)^1 (e^T \uparrow)^2 (e \uparrow)^1 (e \downarrow)^2 (\underline{L}_e \uparrow)^2\right\rangle$ | $H_{13,13} = -\frac{25}{6}B - \frac{5}{3}C + 2\Delta + U$ |

$$H_{1,2} = H_{3,6} = H_{4,7} = H_{5,8} = T_{a_1^T}$$

$$H_{1,3} = H_{2,6} = H_{3,9} = H_{4,10} = H_{5,11} = \sqrt{2}T_{e^T}$$

$$H_{1,4} = H_{2,7} = H_{3,10} = H_{5,12} = T_e$$

$$H_{1,5} = H_{2,8} = H_{3,11} = H_{4,12} = H_{5,13} = \sqrt{2}T_e$$

**TABLE II.** Electronic structure parameters for MnAs nanoclusters embedded in GaAs. $\Delta$, $U$, $(pd\sigma)$ are given in units of eV.

| | nominal oxidation state | coordination number | geometry | crystal system | $\Delta$ | $U$ | $pd\sigma$ | $N^e_{\text{Mn }3d}$ | $N^h_{\text{As }4p}$ |
|---|---|---|---|---|---|---|---|---|---|
| GaAs:MnAs | Mn$^{3+}$ | 6 | distorted octahedral | hexagonal | −2.1 | 3.6 | −0.97 | 5.22 | 1.22 |

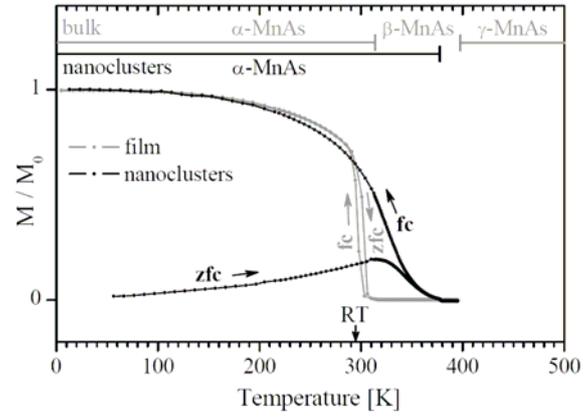

**Figure 1**

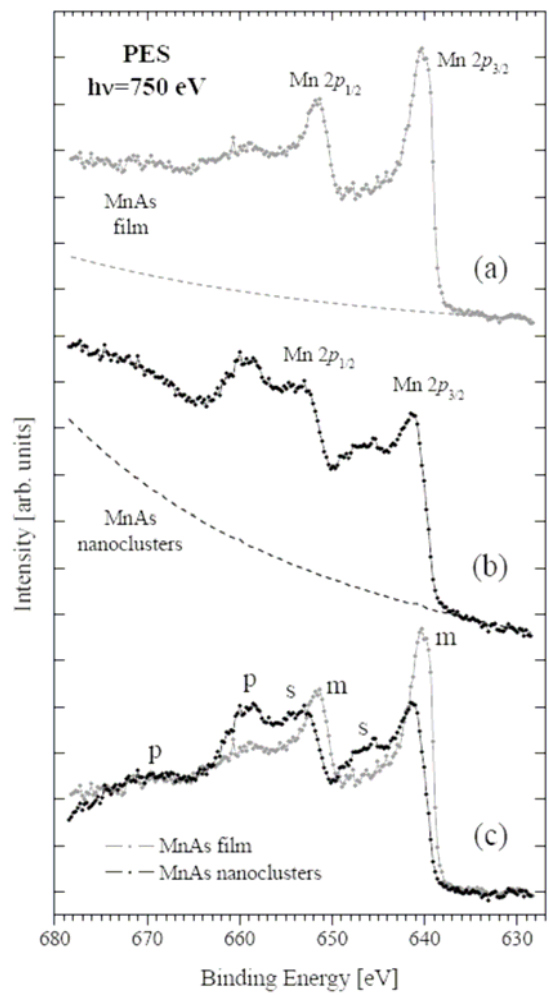

**Figure 2**

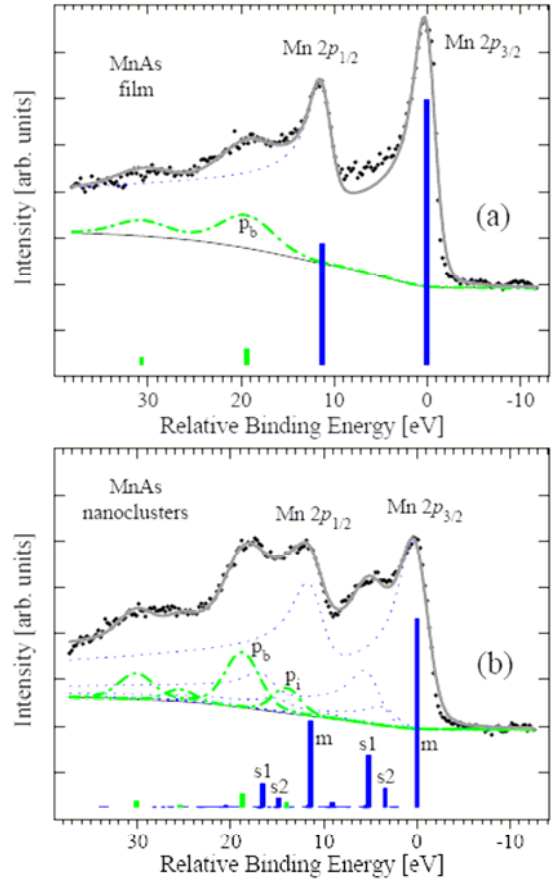

**Figure 3**

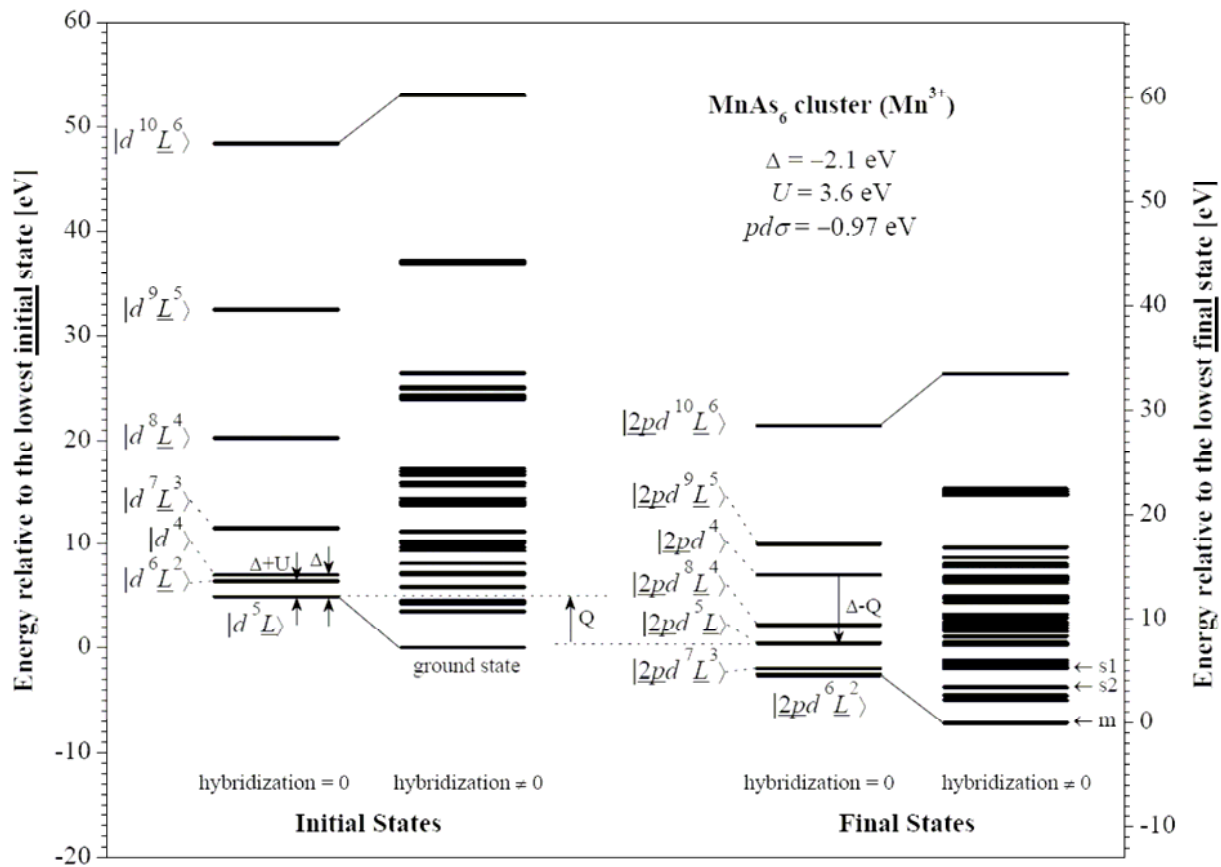

**Figure 4**